# Measurement of electron beam induced sample heating in SEM experiments

Christina Koenig[1], Alice Bastos da Silva Fanta[1,*], Joerg R. Jinschek[1,*]

[1] National Centre for Nano Fabrication and Characterization (DTU Nanolab), Technical University of Denmark (DTU), Kgs Lyngby, Denmark

* Corresponding author: absf@dtu.dk, jojin@dtu.dk

## Abstract

Scanning Electron Microscopy (SEM) experiments provide detailed insights into material microstructures, enabling high-resolution imaging as well as crystallographic analysis through advanced techniques like Electron Backscatter Diffraction (EBSD). However, the interaction of the high-energy electron beam with the material can lead to localized heating, which may significantly impact specimen integrity, especially in applications requiring prolonged beam exposure, for instance when mapping the crystal structure using EBSD. This study examines electron-beam-induced heating effects on a model metal sample (iron), directly measuring the locally deposited electron beam energy with a MEMS-based heating device and validating these measurements through simulations, including Monte Carlo and Finite Element methods. The analysis focuses on the effects of various experimental parameters such as acceleration voltage (from 5 to 30 kV), beam current (from 0.17 nA to 22 nA), dwell time (from 1μs to 1ms) and sample tilt (0° to 70°). The findings reveal that local sample temperatures can increase by up to 70 °C during EBSD experiments, primarily affected by the choice in beam current and acceleration voltage, with beam current having the most significant impact.

*Keywords: SEM, EBSD, beam induced heating, MEMS based heating*

## Highlights

1. MEMS-heating devices enable measurement of electron beam-induced sample heating in SEM.
2. Systematic variation of SEM parameters reveals beam current as the primary factor influencing sample heating.
3. Real-time feedback loops in MEMS-heater enable accurate temperature control during in-situ heating experiments.

## Introduction

The use of a high-energy focused electron beams in a materials characterization experiments using a scanning electron microscope (SEM) is crucial for achieving high-resolution in imaging and analysis but can also affect a specimen's integrity up to the point that the material's microstructure is altered [1, 2, 3]. Such effects, restricted by the electron-beam / sample interaction volume, varies depending on the SEM experimental parameters, including the chosen acceleration voltage and current of the electron beam [4, 5]. Minimizing beam-induced effects is especially important for analytical techniques like electron backscatter diffraction (EBSD), which require prolonged electron beam exposure per pixel. Higher beam currents and longer exposure times improve the signal-to-background ratio and the EBSD pattern quality by allowing more electrons to be collected [6, 7]. However, these conditions also increase the risk of electron beam-induced damage to the sample [2]. In fact, prolonged electron beam exposure during EBSD experiments has been shown to cause unintended structural changes in materials structure. For example, Yi et al. [8] have shown the formation of a graphene film on a 90Cu10Ni alloy after electron beam irradiation during EBSD measurements with 25 kV at 13 nA and an exposure time of 0.8 µs per pixel [8]. Wisniewski et al. [9] investigated pattern degradation during EBSD mapping of annealed glass using step sizes of 1 µm and 0.2 µm. By comparing the average pattern quality of the first scanned line to subsequent ones, they found a 56.6 % loss for the 0.2 µm step and 27.7 % for the 1 µm step, respectively. The pattern degradation, more pronounced at smaller step sizes, was attributed to either crystal lattice damage or overlapping electron beam heating zones [9]. Additionally, it has been shown that the electron beam current during in-situ EBSD tensile testing affects martensitic transformation in 304L steel, with higher currents (32 nA) suppressing the phase instability in the near-surface layer, while a lower beam current (8 nA) showed comparable results to bulk measurements [3].

Egerton et al. categorize the different types of electron-induced damage based on the scattering mechanisms involved. Under standard SEM acquisition conditions with bulk samples, these electron beam / sample interaction mechanisms include electrostatic charging caused by elastic scattering, as well as hydrocarbon contamination and specimen heating resulting from inelastic scattering events [2]. During inelastic scattering, the specimen absorbs a significant portion of the primary electron energy, converting it into phonons or heat within the material [10].

Several studies have attempted to measure the degree of beam-induced heating in SEM [4, 11, 12, 13, 14, 15]. For instance, Holmes et al. [4] used K-type thermocouples as samples inside the chamber to detect temperature changes. They observed a linear rise in temperature as beam current, magnification, and acceleration voltage increased, with a peak temperature of 64 °C at 45 nA and 30 kV [4]. Tokunaga et al. [12] used an infrared (IR) camera positioned behind a germanium window to study the temperature distribution over a sample under electron beam exposure. At beam parameters of 30 kV and 30 nA, the graphite sample reached around 25 °C, while the temperature of the iron membrane sample was measured to reach 30 °C. The differences became more pronounced at even higher beam currents of up to 300 nA, where the temperature of the graphite sample reached approximately 70 °C, while the iron membrane sample exhibited temperatures exceeding 140 °C. The study found that variations in thermal conductivity significantly affected the temperature distribution [12]. However, the IR camera's limited resolution (320x240 pixels, mounted outside the SEM chamber), could limit the detection of localized temperature peaks near the electron beam-irradiated region, reducing the precision of the temperature measurements.

Microelectromechanical systems (MEMS) based sample stages have become a powerful tool for in-situ microscopy, particularly in heating studies, due to their precise temperature control and high thermal stability [16]. Initially developed for transmission electron microscopy (TEM) [16, 17], MEMS-based heaters have also proven to be suitable for in-situ SEM investigations [18, 19, 20]. These heaters operate on the principle of Joule heating, where an electrical current flows through a resistive material, generating heat. A key feature of MEMS-based heaters is the change in resistivity of the heating element with increasing temperature. This defined change in resistivity is pre-calibrated, enabling direct real-time temperature estimation and control through a 4-point probe feedback loop [21].

This study explores electron-beam-induced heating effects in materials under standard SEM acquisition conditions, specifically in EBSD experiments on bulk samples. Using a MEMS-based heating device, we measure sample temperatures during electron beam exposure while systematically varying SEM parameters such as acceleration voltage, beam current, dwell time, and sample tilt. Our experimental findings are supported by Monte Carlo and finite element method (FEM) simulations, which model thermal effects based on the electron beam's interaction volume in the material. By combining both experiments and simulations, we aim to gain deeper insights into the actual local sample temperature during prolonged electron beam exposure in SEM-EBSD mapping experiments.

**Methods**

*Experimental*

The material investigated in this study is Armco-iron, with a purity of 99.85 %. The material was mechanically polished up to a diamond size of 0.25 µm followed by a planar sample lift-out using a ThermoFisher Helios 5 Hydra UX PFIB system. The PFIB sample lift-out workflow is schematically represented in **Figure 1**.

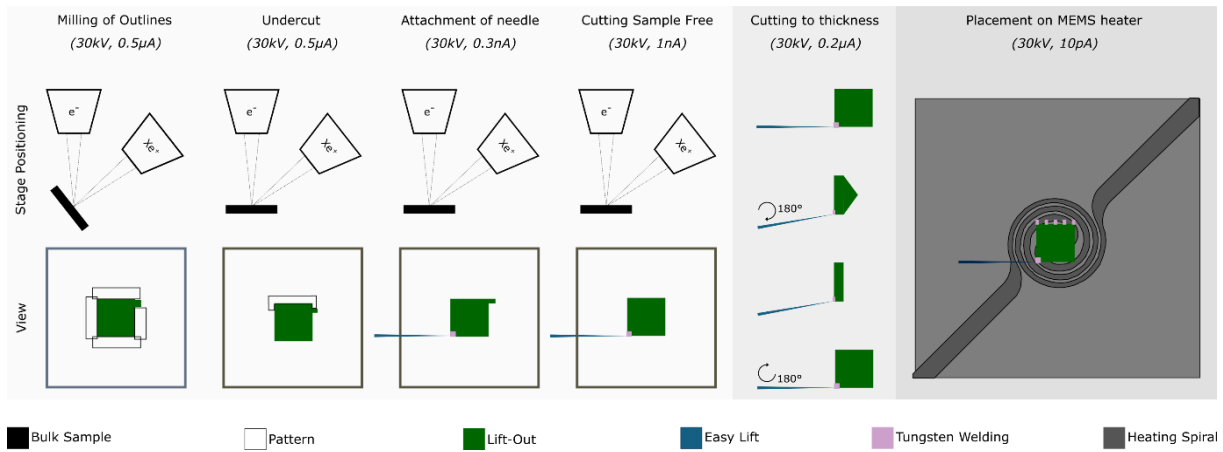

*Figure 1* Schematic representation of workflow for planar lift-out of bulk-like sample and placement of the sample on the surface of the MEMS heating device, indicated are Ion beam currents used for each step

A sample dimensions of 50 µm × 50 µm × 10 µm were chosen to match the size of the central heating area of the MEMS heater, thereby minimizing variations of the local surface temperature in the heating experiment. The sample thickness of 10 µm was specifically chosen so that the sample thickness exceeds the depth of the electron beam interaction volume for 30 kV, see **Figure 3a**.

Using Xenon plasma ions, after an initial bulk-milling (30 kV & 0.5 µA) the sample was thinned down to the desired thickness of 10 µm using 30 kV & 0.2 µA. The sample was then directly placed on a DENS Solutions Wildfire MEMS heating device with silicon nitride (SiN) windows. The sample was attached to the chip surface by tungsten deposition (30 kV, 0.3 nA) using the Xenon ions.

Following the sample lift-out procedure, the MEMS heating device with the sample attached was transferred to a ThermoFisher Helios NanoLab 600 and mounted on a homebuilt heating-stage [22]. This system is equipped to connect the MEMS heater to the *DENS Digiheater* software, which features active feedback control with a fixed interval of 0.3 seconds, with an uncertainty in temperature measurement of ±5%, as specified by the vendor [21].

The MEMS heater was calibrated at a room temperature of 23 °C, which was also set as the experimental baseline temperature. At this setting, the heater operates at its minimum current, supplying just enough power to maintain 23 °C. During the experiment, the MEMS control system continuously monitors the heater's resistivity, converting it to temperature values using a pre-established calibration curve provided by the vendor. When the electron beam interacts with the sample, it introduces additional heat into the system. Since the heater is already operating at its minimum current, it cannot reduce the power further to counteract the excess heat. As a result, the added heat from the electron beam causes a measurable increase in resistivity, which can then be converted into an equivalent temperature increase.

To account for the wide range of microscope parameters normally used in SEM imaging and EBSD experiments, depending on the material being examined, the chosen parameters in our experiments are shown in **Table 1**.

*Table 1 Overview of investigated SEM experiment parameters*

| SEM experiment parameter | Value(s) [Unit] |
|---|---|
| Acceleration Voltage | 5, 10, 20, 30 [kV] |
| Beam Current | 0.17, 1.4, 11, 22 [nA] |
| Dwell time | 0.001, 1 [ms] |
| Sample tilt | 0, 70 [°] |

For each parameter combination, a 10 µm × 10 µm area at the center of the sample was scanned for 30 seconds using the electron beam at a step size of 0.125 µm. Resistivity and therefore temperature readings were continuously recorded throughout the scan and then averaged over the duration to account for minor fluctuations in the measurements.

*Simulations*

Simulations were conducted using the Monte Carlo simulations of electron trajectory in solids software (CASINO V2.4), which models the electron beam as a Gaussian profile. This approach enabled the calculation of electron energy distribution within the selected material under various SEM parameters, including electron beam energy [kV], number of electrons, and beam diameter [23]. In this study, we simulated 10,000 electrons for acceleration voltages of 5, 10, 20, and 30 kV, respectively, assuming a 27.2 nm beam diameter as determined by the SIRAF algorithm [24] (see details in **SI.1**).

Monte Carlo simulations primarily focus on electron trajectories, modeling backscattered electrons, X-ray emissions, and energy absorption within the sample, but a similar Gaussian approximation of the electron beam can be implemented within COMSOL Multiphysics [25]. This enables finite element method (FEM) simulations based on heat transfer in solids to estimate heat generation due to the electron beam exposure. The Gaussian function used to describe the energy intensity profile on the sample surface is defined by the following equation [26]:

$$I(x, y) = \frac{2(V*I)}{\pi r_{\text{Spot}}} \exp\left(\frac{2((x-x_{\text{focus}})^2+((y-y_{\text{focus}})^2)}{r_{\text{Spot}}^2}\right) \quad (1)$$

| | | | |
|---|---|---|---|
| $I(x,y)$ | *Intensity* | $R_{spot}$ | *Electron beam radius* |
| $V$ | *Acceleration Voltage* | $x, y_{focus}$ | *Focus on sample surface* |
| $I$ | *Beam Current* | | |

The simulations were performed using a tetrahedral mesh with a maximum mesh size of 0.5 μm. The initial sample temperature was set at 20 °C prior to electron beam exposure. Material properties were defined using iron from the COMSOL database, with an emissivity of 0.45 [27]. Given that the experimental conditions within a SEM operate under vacuum, only radiation and conduction heat transfer were considered.

For in-situ SEM heating experiments, accurately knowledge of the sample temperature during electron beam exposure is crucial. Therefore, we mimic such an experiment by setting the temperature of the MEMS heater to 200 °C. We then expose the sample on the heater to the electron beam, allowing us to observe how the feedback-loop system adjusts as the electron beam being switched on and off.

**Results and Discussion**

In this study, temperature changes induced by electron beam exposure were systematically measured using a MEMS heating device. Throughout the experiment, resistivity in the MEMS heater is measured and converted to temperature. The average resistivity and temperature over 30 seconds of beam exposure are illustrated in **Figure 2**, shown as a function of acceleration voltage and sample tilt at 0° and 70°.

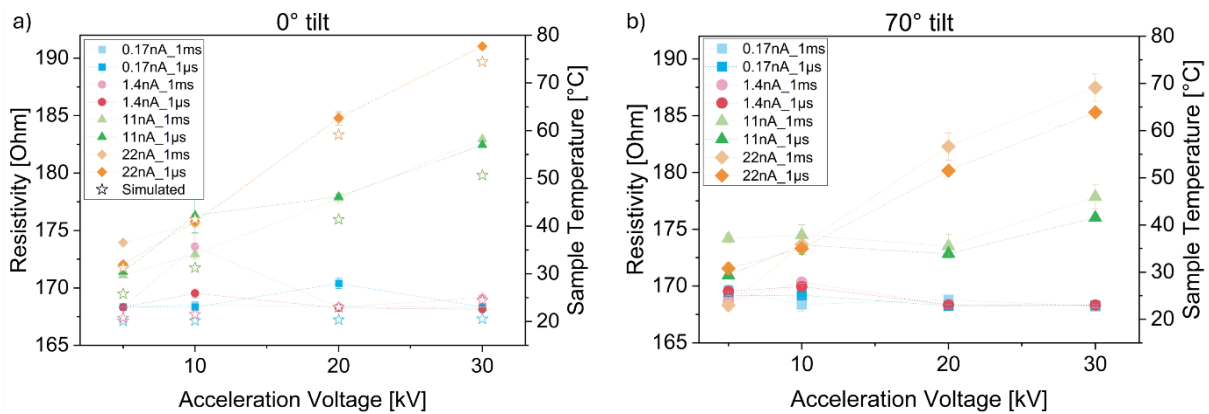

*Figure 2 Measured resistivity and corresponding sample temperature of iron exposed to electron beam at a) 0° tilt and b) 70° tilt, as a function of acceleration voltage, beam current, and dwell time. Data points are grouped by shape according to beam current, with lighter colors representing a dwell time of 1ms and darker colors indicating a dwell time of 1μs.*

The results of our study indicate that both beam current and acceleration voltage significantly influence the temperature rise in the sample, whereas the effect of the beam current is slightly greater. This is evident from the ~16 °C increase in temperature when the beam current is doubled from 11 nA to 22 nA at a constant acceleration voltage of 20 kV, compared to a ~10 °C rise when the acceleration voltage is increased from 20 kV to 30 kV at 11 nA (see **Figure 2a**).

Both beam current and acceleration voltage were expected to significantly influence the sample's temperature as they have a significant effect on electron-matter interactions. The number of scattering events is proportional to the beam current, which is the number of electrons per second that impact the sample [5]. As more electrons interact with the sample, an increased number of inelastic scattering events occur, depositing more energy into the sample. Consequently, higher beam currents lead to an increased deposited energy, resulting in a higher temperature rise. Furthermore, acceleration voltage primarily affects how deeply the electron beam penetrates the sample before losing its entire energy (= depth of the interaction volume). The combined effects of these parameters determine how much energy is absorbed by the sample and how it is spatially distributed, leading to an overall increase in sample temperature.

This spatial energy distribution becomes particularly relevant when considering the tilt angle of the sample. For example, under the same conditions of 30 kV acceleration voltage, 22 nA beam current, and 1 µs dwell time, the increase in sample temperature was measured to be approximately 77 °C at a 0° sample tilt, while a lower temperature increase to 63 °C was measured at a 70° tilt. This difference can be explained by the geometry of the interaction volume. Tilting the sample effectively reduces the interaction volume by partially "opening" it at the surface, reducing the energy absorbed and thus lowering the temperature increase [7].

The effect of dwell time on temperature appears minimal, with shorter dwell times tending to result in slightly lower temperature (see **Figure 2b**). For example, at a sample tilt of 70°, with an accelerating voltage of 20 kV and a beam current of 22 nA, the temperature increases from 51 °C at a dwell time of 1 µs to approximately 57 °C at a dwell time of 1 ms. At a lower beam current of 11 nA, the temperature only increases by around 3 °C, from 33 °C at a dwell time of 1 µs to 36 °C at a dwell time of 1 ms. However, the standard deviations for these last two measurements are overlapping, indicating insignificant differences in temperature between high and low dwell times.

The measured electron beam-induced sample heating in this study shows reasonable agreement with the temperatures measured by Holmes et al. [4], although our measurements seem to provide higher values [4]. For example, while they reported a temperature increase of ~ 46 °C at 30 kV and 23 nA, our experiments recorded increases in sample temperature exceeding 70 °C for the same electron beam parameters. A possible explanation for this discrepancy could be related to the issue raised by Tokunaga et al. [12], who suggested that high conductive heat transfer within the thermocouples used by Holmes et al. may limit measurement accuracy [12].

Tokunaga et al. [12] further reported an average sample temperature increase of 30 °C for 30 kV and 30 nA on an iron membrane sample, whereas our study measured values up to 74 °C in increased sample temperature. Two possible explanations could account for these differences. First, the thickness of the iron membrane used by Tokunaga et al. [12] is only 200 nm, which means that at these electron beam parameters the depth of the interaction volume exceed the sample thickness, potentially altering measurement accuracy (compare with **Figure 3a**). Additionally, based on Figures 2 and 4 in [12] the chosen thermal IR camera setup may have limited spatial resolution, potentially missing localized temperature hot spots from the focused electron beam.

Despite these discrepancies, our results align with both studies in observing a significant increase in sample temperature with higher beam currents and accelerating voltages. This suggests that for EBSD measurements on beam-sensitive samples, if beam-induced heating is the primary damage mechanism, increasing exposure time rather than a higher beam current may be an alternative approach to avoid sample damage while improving the signal-to-background ratio. Our experimental data indicate that this approach is particularly effective for high thermal conductivity materials, where heat dissipates efficiently, minimizing the risk of localized hot spots. However, for insulating materials with limited heat dissipation, longer exposure times may lead to such a hot spot formation, which in turn could result in increased beam-induced damage. Therefore, further confirmation across multiple material systems is needed.

To support our experimental measurements, Monte Carlo and FEM simulations were conducted to analyze electron beam interactions and temperature effects in the sample. **Figure 3a** illustrates the 'energy by position' intensity profile for electron trajectories in iron, showing electron beam directions both perpendicular and parallel to the sample surface at an acceleration voltage of 30 keV. The simulations are based on the assumption of a Gaussian-shaped electron beam profile, which we also assume in the FEM simulations. To demonstrate the similarities, we compare the line profile along the sample surface from Monte Carlo (as indicated by green line) with the FEM's Gaussian profile (30kV, 22nA), as illustrated in **Figure 3b**. The normalized planar profile from our Monte Carlo simulation matches closely the Gaussian approximation used in FEM.

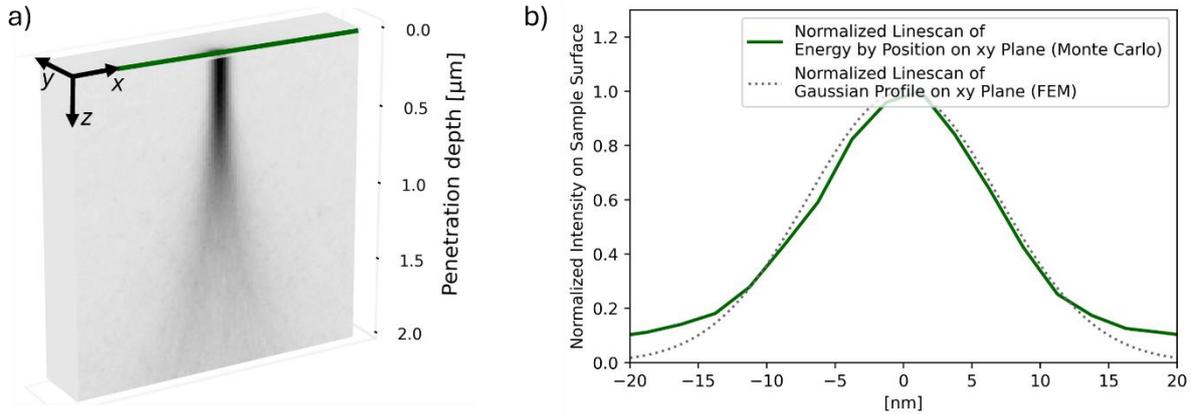

*Figure 3 a) 3D representation of simulated interaction volume from Monte Carlo simulations, visualized as energy by position and b) normalized linescan comparison of the energy distribution on the sample surface between Monte Carlo and FEM simulations*

FEM simulations, shown in **Figure 4**, are used to support experimental observations of the effects of electron beam exposure, specifically examining how variations in acceleration voltage and beam current influence temperature changes within the iron sample.

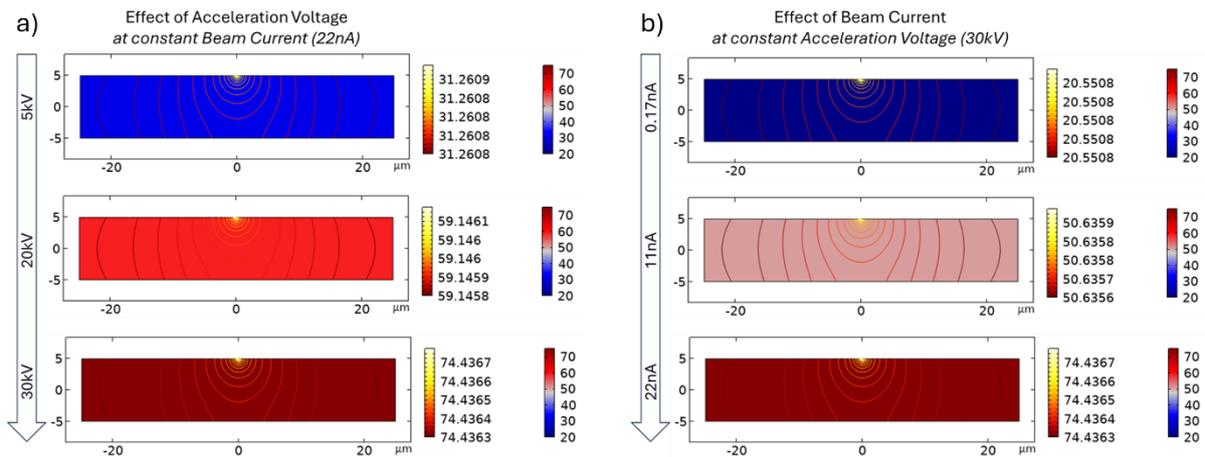

*Figure 4 FEM simulation of temperature variations in the sample after electron beam exposure at 0° tilt. Left color scale corresponds to local temperature distribution; right color scale indicates overall temperature rise with increasing acceleration voltage (a) or beam current (b).*

Focusing on the temperature distribution within the sample for a single electron beam parameter set, FEM simulations indicate only minimal temperature variations within an iron sample. For example, simulations using 30 kV and 22 nA as parameters show only a slight temperature deviation of 0.0004 °C—from 74.4367 °C at the beam center to 74.4363 °C at the edge, see **Figure 4**. This is shown by the contour lines, which highlight a temperature peak at the point where the electron beam is focused. This very small variation can be explained by the sample's high thermal conductivity of 76.2 W/m·K [25] and cannot be detected by current experimental setups. However, when considering the overall temperature evolution, simulations at a constant beam current of 22 nA show that increasing the acceleration voltage from 5 kV to 30 kV leads to a significant temperature rise, reaching a maximum of 74.4 °C at 30 kV (see **Figure 4a**). Similarly, when the acceleration voltage is kept constant at 30kV (see **Figure 4b**), varying the beam current from 0.17 nA to 22 nA results in simulated sample temperatures of 20.5 °C (at 0.17 nA), 50.6 °C (at 11 nA), and 74.4 °C (at 22 nA), respectively. These values align with the measured sample temperatures shown in **Figure 2a**, where FEM simulations closely match the experimental data.

As demonstrated, the choice of electron beam parameters directly impacts the sample's temperature, indicating that localized heating effects are likely, particularly during EBSD measurements. Moreover it highlights the significance of advancements in high-efficiency cameras, such as Direct Electron Detectors, allowing a substantial reduction in beam current without compromising data quality [28].

Our simulations highlight the importance of high thermal conductivity in ensuring uniform heat dissipation, as a critical factor in the MEMS-based approach used to measure beam-induced heat. This enables to measure temperatures near the electron beam focus area, within 25 µm (half the sample width) and 10 µm in depth. While small temperature variations exist, they remain within measurement uncertainties. However, even with this localized detection, temperature can only be measured at the 'backside' of the sample, at the contact point with the MEMS heater, meaning that heat must dissipate first through the sample before detection. This approach to measuring beam-induced heating may face limitations with materials of lower thermal conductivity, where heat transfer to the backside heater would be less efficient. In addition to these considerations, understanding electron beam-induced heating is especially important in in-situ heating studies, where it can alter the set experimental conditions. Leveraging the temperature-sensitive resistivity of the MEMS heater allows for continuous monitoring and real-time adjustment of its temperature by fine-tuning the heater power to maintain the desired experimental conditions. **Figure 5** shows an example where the temperature of a MEMS heater is set to 200 °C and the electron beam is turned on and off.

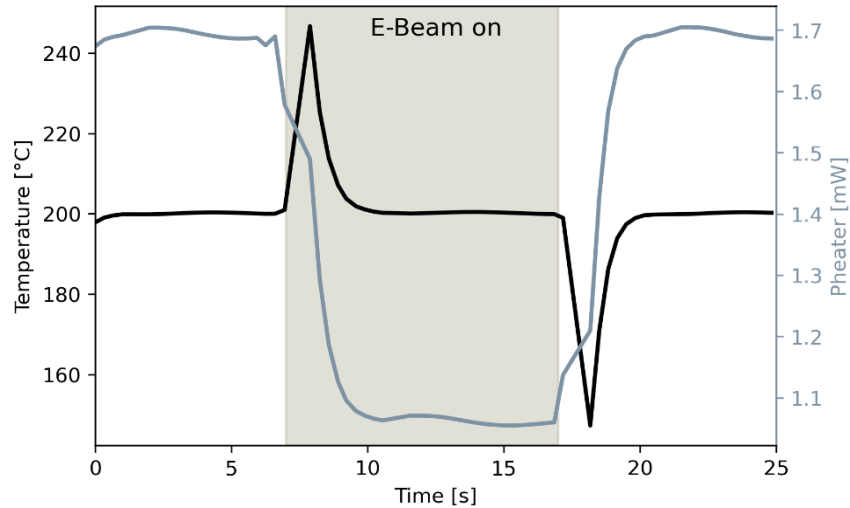

*Figure 5* Response of MEMS heater to electron beam exposure during an in-situ heating experiment with MEMS heater is set to 200°C and the electron beam parameters are chosen to be 30kV acceleration voltage and 22nA beam current.

Initially, the beam is turned off, and the feedback system of the MEMS heater logs a constant power and temperature. When the electron beam is activated, a temperature spike of ~ 46 °C occurs from 200 °C to 246 °C (see **Figure 5**), which the system regulates by reducing the power within a time frame of ~2.5 seconds. This power reduction corresponds to ~0.6 mW, which matches the additional power induced by the electron beam at the selected settings of 30 kV and 22 nA, equaling 0.66 mW. Due to its PID (proportional-integral-derivative) controller, the MEMS heater takes a few seconds to adjust and correct for the additional heat input, eventually reaching steady parameters again after approximately 2.5 seconds. The opposite effect occurs when the beam is turned off. Although the MEMS device can correct for the now "switched off" sample heating by the electron beam (corresponding to a temperature drop by ~ 51 °C), there is a short time delay of ~ 2 seconds when the actual experimental conditions deviate from the controlled settings.

Especially when investigating thermally activated processes through in-situ heating experiments—such as diffusion processes or phase transformations in metals—it is important to consider short-term temperature spikes caused by the electron beam. This is crucial for accurately interpreting results and recognizing possible deviations between actual experimental conditions and set parameters. For materials with low thermal conductivity, we expect that the heater's ability to respond to additional heat will be even slower (i.e. longer delay time) because of delayed and/or limited heat dissipating through the material itself, leading to undetected localized temperature fluctuations or inaccurate measurements, increasing the overall inaccuracy in knowing the SEM sample's surface temperature in such experiments.

## Conclusion

This study successfully demonstrated that electron-beam-induced heating can be effectively measured using MEMS-based heating devices as a sensor, providing insights about localized temperature effects and variations during SEM experiments. Our results identify beam current as the most significant parameter influencing sample temperature, followed by acceleration voltage, sample tilt, and dwell time, with FEM simulations supporting these experimental observations. While the high thermal conductivity of the iron sample enabled efficient heat dissipation, MEMS-based heaters may be less effective as sensors in materials with lower thermal conductivity, highlighting a limitation for estimating beam-induced temperatures in such cases. However, our study shows that combining Monte Carlo simulations with FEM modeling provides an approach to accurately predict temperature variations due to electron-beam-induced heating, making it a versatile tool for analyzing these effects across materials with varying thermal properties.

In EBSD experiments, where electron-beam-induced heating is unavoidable, our findings highlight the need of optimizing SEM microscope parameters to limit potential beam-induced heat damage. In the context of in-situ heating studies, the MEMS heater's capability to adjust in real-time to additional electron beam heating is essential for maintaining accurate control over experimental conditions. However, a considerable response time (i.e. delay time) must be considered. Our findings demonstrate that MEMS devices enable the investigation of thermal processes with reduced beam-induced heating effects.

## Acknowledgment

The authors acknowledge the continued support of their colleagues at the National Centre for Nano Fabrication and Characterization in Denmark (DTU Nanolab) at the Technical University of Denmark (DTU) for their scientific contributions and the many fruitful discussions. Financial support from DTU, enabling a DTU Alliance project in collaboration with Technical University of Munich, is gratefully acknowledged, with special thanks to Prof. Peter Mayr (Chair of Materials Engineering of Additive Manufacturing at the TUM School of Engineering and Design) for his valuable support.

## Declaration of competing interest

The authors declare that they have no known competing financial interests or personal relationships that could have appeared to influence the work reported in this paper.

**Author Contributions**

CK, ABF and JJ contributed to the conceptualization of the study. Data curation and formal analysis were carried out by CK. Funding acquisition was handled by JJ. CK prepared the original draft of the manuscript, while all authors (CK, ABF, and JJ) participated in review and editing.

**Data availability**

The data that support the findings of this study are available from the corresponding author upon reasonable request.

## SI.1 – SEM spot size estimation through SIRAF algorithm

The SIRAF algorithm, as described by Brostrøm et al. [24], was applied to estimate the SEM spot size using a Fourier analysis-based method to determine spatial resolution directly from single images. Gold particles were analyzed at 30 kV and 22 nA, as their non-regular features enable reliable resolution assessment in Fourier space. Here, it is assumed that the resolution limit corresponds to the spot size, which was determined to be 27.2 nm. The steps of the algorithm leading to this measurement are illustrated **Figure SI 1**.

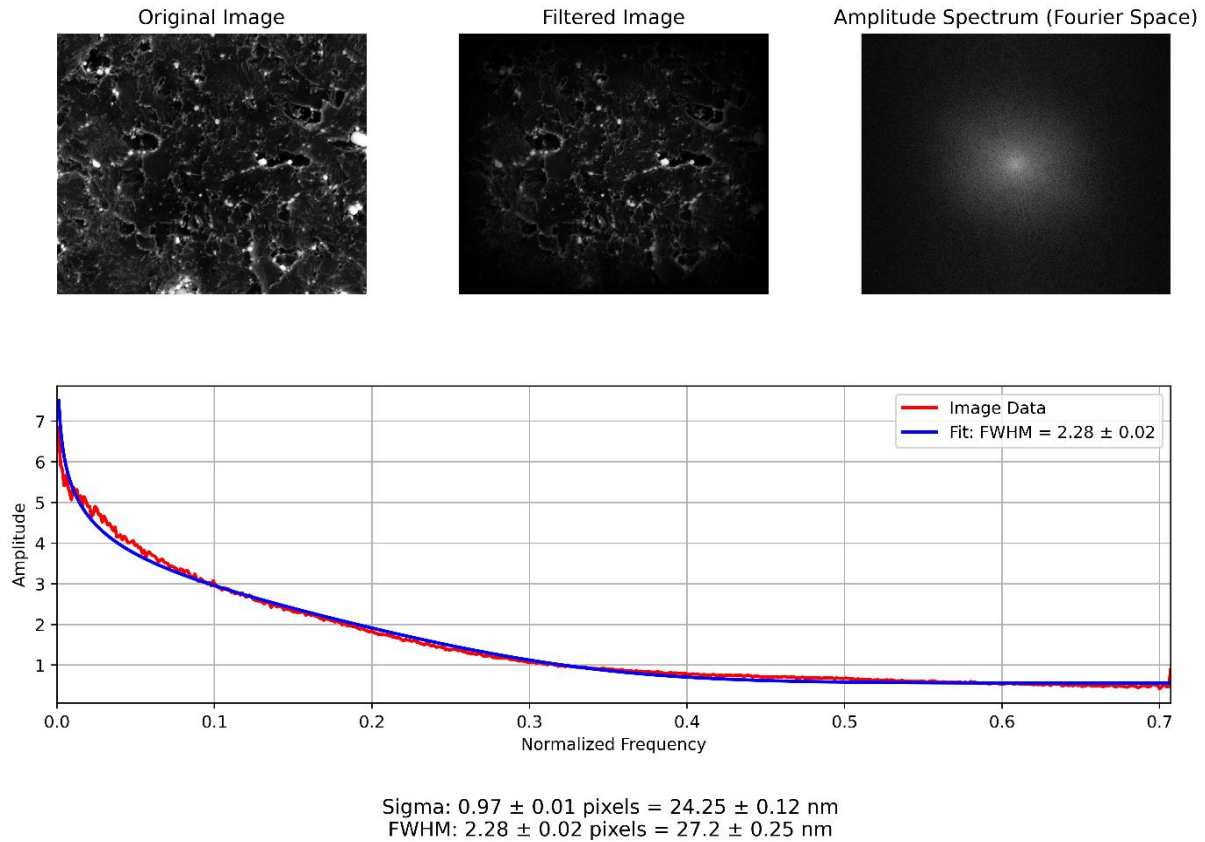

*Figure SI 1 Results of SIRAF algorithm, applied on gold particles, images acquired at 30kV and 22nA with a working distance of 10mm, corresponding to the experimental values.*